\documentclass[aps,prl,reprint,superscriptaddress,longbibliography]{revtex4-1}

\usepackage{amsmath}
\usepackage{mathtools}

\newcommand{\real}[1]{\Re\left\{ #1 \right\}}
\newcommand{\imaginary}[1]{\Im\left\{ #1 \right\}}
\renewcommand{\vec}[1]{\boldsymbol{#1}}
\newcommand{\abs}[1]{\left\vert #1 \right\vert}

\begin{document}

\title{Motion Induced by Light: Photokinetic Effects in the Rayleigh Limit}

\author{David B. Ruffner}
\affiliation{Department of Physics and Center for Soft Matter Research,
New York University, New York, NY 10003}
\affiliation{Spheryx, Inc., New York, NY 10016}
\author{Aaron Yevick}
\author{David G. Grier}
\affiliation{Department of Physics and Center for Soft Matter Research,
New York University, New York, NY 10003}


\begin{abstract}
Structured beams of light can move
small objects in surprising ways.
Particularly striking examples include observations
of polarization-dependent forces acting on optically isotropic
objects and tractor beams that can pull objects
opposite to the direction of the light's propagation.
Here we develop a theoretical framework in which these effects vanish at the
leading order of light scattering theory.
Exotic optical forces emerge instead from interference
between different orders of multipole
scattering.
These effects
create a rich variety of ways to manipulate small objects
with light, so-called photokinetic effects.
Applying this formalism to the particular case of Bessel
beams offers useful insights into the nature of tractor beams
and the interplay between spin and orbital angular momentum in
vector beams of light, including a manifestation of
orbital-to-spin conversion.
\end{abstract}

\maketitle

Beams of light exert forces and torques on illuminated objects.
The resulting photokinetic effects have attracted considerable
interest because of their role in optical micromanipulation \cite{grier03}.
Gradients in the light's intensity give rise to induced-dipole
forces that are responsible for optical trapping \cite{ashkin86}.
Scattering and absorption give rise to
radiation pressure through transfer of the light's linear momentum density.
The momentum density, in turn,
is steered by gradients in the wavefronts' phase \cite{roichman08}
and also by the curl of the light's spin angular momentum density
\cite{berry09,albaladejo09,ruffner12}.
The repertoire of effects governed by the amplitude,
phase and polarization profiles of a structured beam of light
features several surprises.
Circularly polarized beams of light, for example, influence
optically isotropic objects in ways that linearly polarized light cannot
\cite{zhao07,*marrucci11,ruffner12}.
Appropriately structured beams of light can even transport small 
objects upstream
against the direction of propagation
\cite{marston06,lee10,chen11,novitsky11,*wang13,brzobohaty13},
and 
are known as tractor beams by reference to the 
science fiction trope \cite{smith31}.

Here we present a theory for photokinetic effects in vector
beams of light that
lends itself to interpretation of experimental results.
It naturally distinguishes strong and easily observed first-order effects
from more subtle second-order effects, and provides a basis for designing
new modes of light for optical micromanipulation.
Our formulation focuses primarily on
objects that are substantially smaller than the wavelength
of light, the so-called Rayleigh regime, for which we
obtain analytical results.
This analysis explains observation of spin-dependent
forces acting on isotropic objects \cite{ruffner12}, sets limits on
propagation-invariant tractor beams \cite{chen11,novitsky11,*wang13,brzobohaty13}, and predicts
the existence of as-yet unobserved effects such as
orbital-to-spin conversion in helical Bessel beams.
The theory is useful therefore for developing and optimizing
optical micromanipulation tools 
such as first-order tractor beams \cite{lee10}
and knotted force fields \cite{shanblatt11,*rodrigo13}.

\section{Description of a beam of light}

The vector potential of a monochromatic beam of light of frequency
$\omega$ may be decomposed into Cartesian components as
\begin{equation}
  \label{eq:vectorpotential}
  \vec{A}(\vec{r},t) 
  = 
  \sum_{j=0}^2 
  a_j(\vec{r}) \, 
  e^{i\varphi_j(\vec{r})} \, 
  e^{-i \omega t}
  \hat{e}_j
\end{equation}
where $a_j(\vec{r})$ is the real-valued amplitude of the beam
along direction $\hat{e}_j$ at position $\vec{r}$ and $\varphi_j(\vec{r})$ is the
associated phase.
This unusual factorization is useful because it expresses
photokinetic effects
in terms of quantities that can be controlled
experimentally.

For vector potentials that satisfy the Coulomb gauge condition,
\begin{equation}
  \label{eq:coulombgauge}
  \nabla \cdot \vec{A}(\vec{r}) = 0,
\end{equation}
the electric and magnetic fields are related to the vector potential by
\begin{subequations}
\begin{align}
  \label{eq:electricfield}
  \vec{E}(\vec{r},t) 
  & = - \partial_t \vec{A}(\vec{r},t) = i \omega \, \vec{A}(\vec{r},t)
  \quad \text{and}
  \\
  \label{eq:magneticfield}
  \vec{B}(\vec{r},t)
  & = \nabla \times \vec{A}(\vec{r},t).
\end{align}
\end{subequations}
The amplitudes of the vector potential's components 
contribute to the light's intensity through
\begin{equation}
  \label{eq:intensity}
  I(\vec{r}) = \frac{1}{2\mu c} \abs{\vec{E}(\vec{r},t)}^2
  = \frac{\omega^2}{2 \mu c} \, \sum_{j=0}^2a_j^2(\vec{r}).
\end{equation}
With this, the local polarization may be written as
\begin{equation}
  \label{eq:polarization}
  \hat{\epsilon}(\vec{r}) = \frac{2 \mu c}{\omega^2I(\vec{r})}\sum_{j=0}^2 a_j(\vec{r}) e^{i
    \varphi_j(\vec{r})} \hat{e}_j.
\end{equation}
This description provides a point of departure
for discussing light's properties and its
ability to exert forces on small objects.

\section{Properties of a beam of light}

The time-averaged linear momentum density carried by the beam of light is given by Poynting's
theorem,
\begin{equation}
  \vec{g}(\vec{r})
  = 
  \frac{1}{2\mu c^2}
  \real{\vec{E}^\ast(\vec{r},t) \times \vec{B}(\vec{r},t)}.
  \label{eq:poynting}
\end{equation}
where $c$ is the speed of light in the medium.
Expressing $\vec{g}(\vec{r})$  
in terms of experimentally controlled quantities,
\begin{equation}
  \vec{g}(\vec{r})
  =
  \frac{\omega}{2\mu c^2} \, \sum_{j=0}^2 a_j^2(\vec{r}) \, \nabla \varphi_j(\vec{r})
  + \frac{1}{2} \nabla \times \vec{s}(\vec{r}),
  \label{eq:momentumdensity}
\end{equation}
reveals that the light's momentum is guided both by phase gradients
\cite{roichman08,ruffner12} and also by the curl of the
time-averaged spin angular momentum density
\cite{ruffner12,albaladejo09,berry09,ruffner13}
\begin{subequations}
  \label{eq:spindensity}
  \begin{align}
    \vec{s}(\vec{r}) 
    & = 
      \frac{\omega}{2\mu c^2}
      \imaginary{
      \vec{A}^\ast(\vec{r},t) \times \vec{A}(\vec{r},t)
      } \\
    & =
      \frac{i}{\omega c} I(\vec{r}) \, 
      \hat{\epsilon}(\vec{r}) \times \hat{\epsilon}^\ast(\vec{r}).
  \end{align}
\end{subequations}

A beam of light
also can carry orbital angular momentum
\cite{humblet43,allen92,*allen99,berry09},
\begin{subequations}
  \label{eq:orbital}
  \begin{align}
    \vec{\ell}(\vec{r}) 
    & = 
      i \frac{\omega}{2 \mu c^2}
      I(\vec{r})
      \sum_{j = 0}^2
      \epsilon_j (\vec{r})
      \left[
      \vec{r} \times \nabla \epsilon_j^\ast(\vec{r})
      \right] \\
    & =
      \frac{\omega^2}{2 \mu c}
      \sum_{j = 0}^2 a_j^2(\vec{r})
      \left[
      \vec{r} \times \nabla \varphi_j(\vec{r}),
      \right],
  \end{align}
\end{subequations}
which is distinct from its spin angular momentum.
Equation~\eqref{eq:orbital} suggests little connection
between $\vec{\ell}(\vec{r})$ and $\vec{s}(\vec{r})$.
A direct connection has been noted, however,
in experimental studies of particles' circulation in
circularly polarized optical traps
\cite{oneil02,zhao07,*marrucci11,ruffner12}
and has been dubbed spin-to-orbital conversion.
The converse process by which the orbital angular
momentum content of a beam imbues the light
with spin angular momentum has received
less attention.

\section{Optical forces in the Rayleigh regime}
\subsection{First-order photokinetic effects}

Optical forces arise from the transfer of momentum from the beam
of light
to objects that scatter or absorb the light.
They are not simply proportional to $\vec{g}(\vec{r})$, however,
but rather arise from more subtle mechanisms.
To illustrate this point, we consider the force exerted by a beam of
light on an object that is smaller than the wavelength of light.
Taking this limit permits us to adopt the Rayleigh approximation that
the light's instantaneous electric field is uniform across the
illuminated object's volume.
The electromagnetic field then induces electric and magnetic dipole moments in
the particle,
\begin{subequations}
\begin{align}
  \label{eq:electricdipolemoment}
  \vec{p}(\vec{r},t) & = \alpha_e \, \vec{E}(\vec{r},t) \quad
  \text{and} \\
  \label{eq:magneticdipolemoment}
  \vec{m}(\vec{r},t) & = \alpha_m \, \vec{B}(\vec{r},t),
\end{align}
\end{subequations}
respectively,
where $\alpha_e = \alpha_e^\prime + i \alpha_e^{\prime\prime}$
is the particle's complex electric polarizability and $\alpha_m$ is
the corresponding magnetic polarizability.

The induced electric dipole moment
responds to the light's electromagnetic fields through the time-averaged
Lorentz force \cite{gordon73,*chaumet00}
\begin{align}
  \label{eq:lorentzforce}
  \vec{F}_e(\vec{r})
  & = 
    \begin{multlined}[t][0.7\displaywidth]
      \frac{1}{2} \,
      \real{ 
        \left(\vec{p}(\vec{r},t) \cdot \nabla \right)
        \vec{E}^\ast(\vec{r},t)} \\
      +
      \frac{1}{2} \,
      \real{
        \partial_t \vec{p}(\vec{r},t)
        \times\vec{B}^\ast(\vec{r},t)},
    \end{multlined} \\
  & =
  \frac{1}{2} \, \real{
    \alpha_e \sum_{j = 0}^2 E_j(\vec{r},t) \, \nabla \,
    E_j^\ast(\vec{r},t)
  }.
\end{align}
Expressed in terms of experimentally controlled parameters,
the electric dipole contribution to the 
time-averaged force is
\begin{equation}
  \label{eq:electricdipoleforce}
  \vec{F}_e(\vec{r}) = 
  \frac{1}{2} \mu c \, \alpha_e^\prime \, \nabla I(\vec{r})
  + \mu c \, \alpha_e^{\prime\prime} \, 
  \sum_{j=0}^2  a_j^2(\vec{r}) \,
  \nabla \varphi_j(\vec{r}).
\end{equation}

The first term in Eq.~\eqref{eq:electricdipoleforce} is the manifestly
conservative intensity-gradient force that is responsible for trapping
by single-beam optical traps \cite{ashkin86}.
The second describes a non-conservative force \cite{roichman08,ruffner13}
that is proportional to the phase-gradient contribution to
$\vec{g}(\vec{r})$.
This is the radiation pressure experienced
by a small 
particle and is responsible for the transfer of orbital angular
momentum from helical modes of light \cite{he95a,*gahagan96,*simpson96}.
Because $\alpha_e^{\prime\prime} > 0$ for scattering by
conventional materials, radiation pressure tends to drive
illuminated objects downstream along the direction of the light's propagation.

Surprisingly, the spin-curl contribution to $\vec{g}(\vec{r})$ has no
counterpart in $\vec{F}_e(\vec{r})$, 
this missing term having been canceled \cite{ruffner13} by the
induced dipole's coupling to the magnetic field in
Eq.~\eqref{eq:lorentzforce}.
This means that the radiation pressure experienced by a small particle
is not simply proportional
to the light's momentum density \cite{albaladejo09,ruffner13},
as might reasonably have been assumed.
The absence of any spin-dependent contribution
may be appreciated because $\vec{s}(\vec{r})$
involves off-diagonal terms in the components of the polarization whereas
$\vec{F}_e(\vec{r})$ does not \cite{ruffner13}.
Compelling experimental evidence for spin-dependent forces acting
on isotropic objects \cite{ruffner12} therefore
cannot be explained as a first-order photokinetic effect.

The corresponding result for magnetic dipole scattering has a
form \cite{chaumet09,*nietovesperinas10} 
analogous to Eq.~\eqref{eq:electricdipoleforce},
\begin{equation}
  \label{eq:magneticdipole}
  \vec{F}_m(\vec{r}) 
  = 
  \frac{1}{2} 
  \real{
    \alpha_m
    \sum_{j = 0}^2
    B_j(\vec{r},t) \nabla B_j^\ast(\vec{r},t)
  }.
\end{equation}
This contribution to the total optical force similarly is comprised of
a conservative intensity-gradient term and a non-conservative
phase-gradient term, with no contribution from the light's
spin angular momentum density.

At the dipole order of light scattering, the photokinetic forces on
optically isotropic scatterers are
governed by intensity gradients and phase gradients, and are
essentially independent of the light's polarization.
Higher multipole moments \cite{chen11}
similarly contribute
to the conservative intensity-gradient force and 
to the radiation pressure.  None of these terms,
however, feature contributions from the light's
spin angular momentum.

Equation~\eqref{eq:electricdipoleforce} also
constrains designs for tractor beams.
A beam propagating along $\hat{z}$ 
with $\vec{g}(\vec{r}) \cdot \hat{z} > 0$ would exert a retrograde
force if $\vec{F}_e(\vec{r}) \cdot \hat{z} < 0$.
In a propagation-invariant beam satisfying
$\partial_z I(\vec{r}) = 0$, we expect
$\sum_{j=0}^2 a_j(\vec{r}) \, \partial_z \varphi_j(\vec{r}) > 0$.
The phase-gradient
term in Eq.~\eqref{eq:electricdipoleforce} therefore has a positive
axial projection for conventional materials
with $\alpha_e^{\prime\prime} > 0$.
This means that propagation-invariant beams 
cannot act as tractor beams for induced electric dipoles.
A similar line of reasoning yields an equivalent result
for $\vec{F}_m(\vec{r})$.
The apparent absence of propagation-invariant tractor beams at
the dipole order of multipole scattering
is consistent with numerical studies \cite{chen11,novitsky11}
that find no pulling forces for particles much smaller
than the wavelength of light.

\subsection{Second-order photokinetic effects}

Exotic effects such as polarization-dependent forces and
tractor beam action
are restored by the interference between the scattered fields.
For the particular case of the electric and magnetic dipole fields,
such interference leads to a force of the form
\cite{chen11,chaumet09,*nietovesperinas10}
\begin{align}
  \vec{F}_{em}(\vec{r})
  & =
  - \frac{k^4}{12 \pi c \epsilon_0}
  \real{
    \vec{p}^\ast(\vec{r},t) \times \vec{m}(\vec{r},t) 
  } \\
  & =
    \frac{k^3 \mu}{12 \pi \epsilon_0} \biggl[
    - \imaginary{\alpha_e^\ast \alpha_m} \nabla I(\vec{r}) \nonumber \\
  & \quad
    + 
    \epsilon_0 \omega^2
    \imaginary{\alpha_e^\ast \alpha_m}
    \real{ \left[\vec{A}^\ast(\vec{r},t) \cdot \nabla\right]
    \vec{A}(\vec{r},t)}  \nonumber \\
  & \quad\quad
    -
    2 \omega \, \real{\alpha_e^\ast \alpha_m}
    \vec{g}(\vec{r}) 
    \biggr]. 
    \label{eq:interferenceforce} 
\end{align}
The first term in Eq.~\eqref{eq:interferenceforce}
contributes to the intensity-gradient trapping force.
The second
describes a non-conservative force that appears not to have been
discussed previously.  It is influenced by
spatial variations in the light's polarization but is symmetric under
exchange of the components' indexes and so does not depend on
the spin angular momentum density.
The third is
proportional to the linear momentum density
and therefore includes both phase-gradient and
spin-curl contributions.

Interference between electric and magnetic dipole scattering
therefore can account for 
spin-dependent forces of the kind observed in optical trapping
experiments \cite{simpson97,oneil02,zhao07,ruffner12}.
It can be shown, moreover, 
that spin-dependent contributions to optical
forces appear at higher orders of multipole scattering due to interference
between the fields scattered by induced electric and magnetic multipoles.
Higher-order effects than those captured by
Eq.~\eqref{eq:interferenceforce} may be accentuated in
particles larger than the wavelength of light, particularly through
Mie resonances.

Equation~\eqref{eq:interferenceforce}
also suggests a mechanism by which propagation-invariant modes
can act as tractor beams.
The first two terms of Eq.~\eqref{eq:interferenceforce}
are directed substantially transverse to
$\vec{g}(\vec{r})$,
and so cannot contribute to retrograde forces.
The prefactor of the third term is negative for conventional
materials, and thus inherently describes a pulling force.
The overall axial force can be directed upstream for objects
that satisfy
\begin{equation}
  \label{eq:condition}
  \real{\alpha_e^\ast \alpha_m} > 
  \frac{\mu^2}{k^2} \,
  \imaginary{c \, \alpha_e + \frac{1}{c}\alpha_m}.
\end{equation}
This condition is not generally met for particles smaller than
the wavelength of light, which is why tractor-beam action
generally is anticipated for larger particles whose Mie resonances
favor forward scattering \cite{chen11,novitsky11,novitsky12}.

\section{Application to Bessel beams}

As an application of this formalism, we consider the forces exerted by a
monochromatic Bessel beam \cite{zhou84,durnin87,*durnin87a,*mcdonald00},
whose vector potential may be written in
cylindrical coordinates, $\vec{r} = (\rho,\theta,z)$, as
\begin{subequations}
\label{eq:besselbeam}
\begin{equation}
  \label{eq:besselbeamvectorpotential}
  \vec{A}_{m,\alpha}(\vec{r},t)
  =
  a_0 e^{-i\omega t} \hat{\vec{P}}
    J_m(\tilde{\rho}) \,
    e^{i m \theta + i \cos\alpha \, k z}.
\end{equation}
Here, $a_0$ is the wave's amplitude, $k = \omega/c$ is the wavenumber of an equivalent
plane wave at frequency $\omega$,
$\tilde{\rho} = \sin\alpha \, k \rho$ is a dimensionless radial coordinate,
and the operator $\hat{\vec{P}}$ projects
the scalar wavefunction into a vector field
satisfying the Coulomb
gauge condition, Eq.~\eqref{eq:coulombgauge}.
The projection operator
\footnote{The associated operator for the transverse magnetic mode
  is \protect\cite{zhou84}
\begin{equation}
  \label{eq:transversemagneticgaugeoperator}
  \hat{\vec{P}}_\phi^\text{TM}
  =
  - \frac{1}{k^2} \, \nabla \times \nabla \times \hat{z}.
\end{equation}
The Bessel beam described by Eq.~(1) in Ref.~[\onlinecite{chen11}]
may be obtained with a linear combination of
$\hat{\vec{P}}_\phi^\text{TE}$ and
$\hat{\vec{P}}_\phi^\text{TM}$.
},
\begin{equation}
  \label{eq:azimuthalgaugeoperator}
  \hat{\vec{P}}_\phi^\text{TE}
  = 
  - \frac{1}{k} \nabla \times \hat{z},
\end{equation}
\end{subequations}
describes
a transverse electric Bessel beam that is azimuthally polarized for 
$m = 0$.
Using operator notation not only yields compact expressions
for the vector Bessel beams,
but also clarifies the symmetries of their wave functions.

Bessel beams are characterized by 
a convergence angle $\alpha$ that reduces the axial
component of the momentum density relative to that of a plane wave.
An object that scatters light into the forward direction
might thereby increase the momentum density in the beam, and so
would have to recoil upstream to conserve momentum.
This is the basis for the proposal \cite{marston06,chen11,novitsky11,*wang13}
that a Bessel beam can act as a tractor beam.

The vector potential in Eq.~\eqref{eq:besselbeam}
also is characterized by an integer winding number, $m$, that 
governs the helical pitch of the beam's wavefronts
and endows the light with orbital angular momentum 
\cite{allen92,*allen99,volkesepulveda02}.
This is expressed in the axial component of the orbital angular momentum density,
\begin{equation}
  \label{eq:bessellz}
  \vec{\ell}(\vec{r}) \cdot \hat{z}
  = m \, a_0^2 \, \frac{\omega}{2 \mu c^2} J_m^2(\tilde{\rho}),
\end{equation}
which is proportional to $m$.

Remarkably, this Bessel beam's spin angular momentum density,
\begin{equation}
  \label{eq:besselsz}
  \vec{s}(\vec{r})
  =
  m \, a_0^2 \, \frac{1}{2 \mu \omega} \frac{1}{\rho} \,
  \frac{d  J_m^2(\tilde{\rho})}{d \rho} \hat{z},
\end{equation}
also is proportional to the helical winding number $m$.
Equation~\eqref{eq:besselsz} therefore describes
orbital-to-spin angular momentum conversion, 
an effect that appears not to have
been described previously.
Having wavefront helicity control the light's degree of spin polarization
complements spin-to-orbital conversion \cite{zhao07,*marrucci11}
in which a beam's state of polarization contributes to its wavefronts' helicity.
Like spin-to-orbital conversion,
orbital-to-spin conversion vanishes in the paraxial limit
($\alpha \to 0$).

In considering the forces exerted by Bessel beams, it is convenient
to replace the dipole polarizabilities by the first-order
Mie scattering coefficients, $a_1$ and $b_1$, which are
dimensionless and tend to be of the same order of magnitude
for small particles.
They are related to the polarizabilities by
$\alpha_e = i 6 \pi \epsilon_0 n_m^2 a_1/k^3$
and
$\alpha_m = i 6 \pi b_1 / (\mu k^3)$
\cite{bohren83,*mishchenko02,chen11}.
With this substitution, the net optical force due to dipole scattering,
$\vec{F}(\vec{r}) = \vec{F}_e(\vec{r}) + \vec{F}_m(\vec{r}) +
\vec{F}_{em}(\vec{r})$, has an axial component
\begin{multline}
  \label{eq:axialbesselforce}
  \frac{\mu F_z(\vec{r})}{3 \pi k^2 a_0^2 \cos\alpha \sin^2\alpha}
  =
  \real{a_1 + \cos^2\alpha \, b_1} f_{m,\alpha}^2(\tilde{\rho}) \\
  + 
  \real{b_1} 
    \sin^2\alpha \, J_m^2(\tilde{\rho})
  -
  \real{a_1^\ast b_1} f_{m,\alpha}^2(\tilde{\rho}),
\end{multline}
where
\begin{equation}
  f_{m,\alpha}^2(\tilde{\rho}) =
  \left[\frac{d J_m(\tilde{\rho})}{d\tilde{\rho}} \right]^2
  +
  \left[ \frac{J_m(\tilde{\rho})}{\tilde{\rho}} \right]^2 .
\end{equation}
The first two terms in Eq.~\eqref{eq:axialbesselforce} arise from
electric and magnetic dipole scattering, respectively, and tend to push the
scatterer along the $+\hat{z}$ direction.
The third term, which arises from interference, generally is negative
and describes a pulling force.
The Bessel beam acts as a tractor beam 
only if this second-order term dominates the first-order terms.
This condition is met by
scatterers whose polarizabilities satisfy Eq.~\eqref{eq:condition}.
Similar results can be obtained for the radially
polarized Bessel beam.
On the basis of these considerations, we conclude that 
a Bessel beam only acts as a tractor beam
for Rayleigh particles under exceptional circumstances.
The formulation of first- and second-order
optical forces 
reveals that this limitation is a generic feature of propagation-invariant
beams of light and is not specific to Bessel beams.

Distinguishing first- and second-order photokinetic effects
is useful for assessing the nature of the forces exerted by beams of light on
Rayleigh particles.
Trapping by optical tweezers and the torque
exerted by transfer of orbital angular momentum are examples of
first-order effects.
Second-order effects include the spin-curl force, which has been
studied experimentally \cite{ruffner12} and observed numerically
\cite{sun08}, but has not previously been explained theoretically
\cite{albaladejo09,ruffner13}. 
The pulling force exerted by propagation-invariant tractor beams,
including Bessel beams, similarly turns out to be a second-order
effect.

Orbital-to-spin conversion similarly emerges as a property of
helical Bessel beams and is the counterpart to spin-to-orbital
conversion, which has been widely discussed 
\cite{oneil02,zhao07,*marrucci11}.
Beyond revealing subtle properties of vector light waves,
this formalism also provides 
a basis for developing new modes of optical micromanipulation.
Whereas tractor beams with continuous propagation invariance
are inherently limited in their ability to transport small objects, 
Eq.~\eqref{eq:electricdipoleforce} provides guidance for designing
more powerful and longer-range \cite{ruffner14}
tractor beams with discrete propagation invariance that
can operate at dipole order.
Solenoidal waves \cite{lee10} appear to be an example of such first-order 
tractor beams.  The theory of photokinetic effects suggests how to
optimize such modes and how to discover new examples.

This work was supported by the National Science Foundation through
Grant Number DMR-1305875.  We are grateful to Giovanni Milione
for valuable discussions.

%

\end{document}